\documentclass[aps,twocolumn,showpacs]{revtex4}
\usepackage{amsmath}
\usepackage{epsfig}
\usepackage{multirow}

\begin{document}

\title{Exploring possible $^3_{\Lambda_c}\text{H}$ bound states through $p\Lambda_c$ femtoscopic correlations}
\author{Ye Yan$^{1}$}
\author{Qi Huang$^2$}
\author{Qian Wu$^{3,4}$}\email{qwu@nju.edu.cn (Corresponding author)}
\author{Hongxia Huang$^2$}\email{hxhuang@njnu.edu.cn(Corresponding author)}
\author{Jialun Ping$^2$}

\affiliation{$^1$Department of Physics, Changzhou University of Information Technology, Changzhou 213164, China}
\affiliation{$^2$Department of Physics, Nanjing Normal University, Nanjing 210023, China}
\affiliation{$^3$School of Physics, Nanjing University, Nanjing 210000, China}
\affiliation{$^4$Institute of Modern Physics, Chinese Academy of Sciences, Lanzhou 730000, China}

\begin{abstract}
	
The femtoscopic correlation technique in relativistic heavy-ion collisions provides a unique opportunity to investigate hadron-hadron interactions and possible exotic states.
In this work, we study the $p\Lambda_c$ correlation function and its sensitivity to the low-energy $N\Lambda_c$ interaction related to possible $^3_{\Lambda_c}\mathrm{H}$ bound states.
Based on the quark delocalization color screening model, three interaction scenarios with different strengths are constructed, and the corresponding spin-averaged $p\Lambda_c$ correlation functions are calculated within the Koonin--Pratt formalism.	
The results demonstrate that the correlation function is sensitive to the $p\Lambda_c$ interaction strength, with coupled-channel effects and $S$-$D$ wave mixing producing additional enhancements in the correlation signal.
These findings suggest that future $p\Lambda_c$ femtoscopic measurements at relativistic heavy-ion collision experiments can provide valuable constraints on the interaction between charmed baryons and nucleons and offer guidance for exploring possible heavy-flavor hypernuclei.
	
\end{abstract}


\maketitle

\setcounter{totalnumber}{5}

\section{Introduction}
\label{sec:introduction}

Femtoscopic correlations in relativistic heavy-ion collisions have emerged as a powerful tool for investigating the properties of hadronic interactions at low energies~\cite{Wiedemann:1999qn,Lisa:2005dd,ExHIC:2017smd,Fabbietti:2020bfg,Liu:2024uxn}. 
By analyzing the momentum correlations between particles produced in the final state, femtoscopy provides direct access to the space-time structure of the particle-emitting source as well as the final-state interactions between hadrons. 
Extensive experimental and theoretical studies, such as those involving nucleon--nucleon~\cite{Koonin:1977fh,Lednicky:1981su,STAR:2015kha,ALICE:2020ibs,ALICE:2025wuy,SRIT:2026qkv,Xi:2026vrp}, nucleon--hyperon~\cite{ALICE:2019hdt,ALICE:2020mfd,ALICE:2021njx,Fu:2024btw,STAR:2005rpl,STAR:2018uho,Ohnishi:2016elb,Morita:2016auo,Hatsuda:2017uxk,Haidenbauer:2021zvr,Garrido:2024pwi}, and hyperon--hyperon~\cite{ALICE:2018ysd,Ohnishi:1998at,Morita:2014kza,Haidenbauer:2018jvl,Morita:2019rph,Ohnishi:2021ger,Kamiya:2021hdb,Liu:2022nec,STAR:2014dcy,ALICE:2022uso,Sarti:2025sdo} systems, have demonstrated the capability of femtoscopy to reveal the nature of strong interactions and search for possible hadronic bound states.
With the continuous development of heavy-ion collision experiments at the Large Hadron Collider (LHC) and the Relativistic Heavy Ion Collider (RHIC), femtoscopic studies involving increasingly diverse hadronic systems have become an important avenue for exploring the strong interaction in the nonperturbative regime.

With the successful applications in nucleon and strange hadron systems, femtoscopic studies have been further extended to hadrons containing heavy quarks.
In particular, the $pD$ correlation has been proposed as one of the pioneering examples of heavy-flavor femtoscopy~\cite{ALICE:2022enj}, demonstrating the potential of this technique to investigate the interactions between nucleons and charm hadrons.
Meanwhile, the interactions involving heavy-flavor hadrons have attracted extensive theoretical interest.
Recently, many studies have investigated systems involving charm mesons~\cite{ALICE:2022enj, Kamiya:2022thy, Liu:2023uly, Vidana:2023olz, Albaladejo:2023pzq, Ikeno:2023ojl, Liu:2023wfo, Torres-Rincon:2023qll, Liu:2023huu, Albaladejo:2023wmv, Li:2024tof, Liu:2024nac, Abreu:2025jqy, Liu:2025oar, Etminan:2025tiy, Barbat:2025orm, Agatao:2025ckp, Liu:2025wwx, Zhang:2025szg, Liu:2025nze, Ge:2026moy, Zhao:2026tpj, Shen:2025qpj, Shi:2026hwn, Song:2026spz}, including $D^{(*)}$, $D_s$, and charmonium, aiming to understand the nature of near-threshold structures and their possible hadronic molecular interpretations, such as the $X(3872)$, $T_{cc}$, and $P_c$ states.
These studies highlight the importance of constraining heavy-flavor hadron interactions in the nonperturbative regime of Quantum Chromodynamics (QCD)~\cite{Barbat:2026drc, Feijoo:2023sfe, Lai:2026gql, Liu:2026zlk, Jia:2026iqo, Ikeno:2025bsx}.
Experimentally, the continuous improvement of heavy-ion collision experiments at the LHC and RHIC has made femtoscopic measurements involving heavy-flavor particles increasingly feasible, providing new opportunities to access these poorly constrained interactions.
These developments motivate the extension of femtoscopic studies from charmed mesons to charmed baryons, such as the $\Lambda_c$ baryon, where the interaction with nucleons remains largely unconstrained.

The $N\Lambda_c $ interaction is one of the most fundamental charm--nucleon interactions and plays an important role in exploring the strong interaction in the charm sector. 
However, the $N \Lambda_c$ interaction remains poorly constrained due to the absence of experimental data. 
Theoretical investigations of the $N \Lambda_c$ interaction have been performed within various approaches, including meson-exchange models, constituent quark models, and lattice QCD.
Phenomenological studies based on meson-exchange and quark models generally predict relatively strong attractive $N \Lambda_c$ interaction~\cite{Liu:2011xc,Maeda:2015hxa,Garcilazo:2019ryw,Vidana:2019amb}, whereas lattice QCD calculations from the HAL QCD Collaboration suggest a considerably weaker attraction~\cite{Miyamoto:2017tjs,Miyamoto:2017ynx}.
Furthermore, nonrelativistic chiral effective field theory (EFT) studies based on the extrapolation of lattice QCD results to the physical pion
mass suggest that the $N \Lambda_c$ interaction is moderately attractive but weaker than some phenomenological predictions~\cite{Haidenbauer:2017dua}.
Meanwhile, covariant chiral EFT analyses show that the interaction can be sensitive to coupled-channel effects and may even become repulsive in the spin-triplet channel~\cite{Song:2020isu}.
Such theoretical uncertainties motivate further studies of the $N \Lambda_c$ interaction.

To better constrain the poorly known $N\Lambda_c$ interaction, theoretical studies of the $p\Lambda_c$ femtoscopic correlation function have been
performed recently~\cite{Haidenbauer:2020kwo,Zheng:2026qtk}. 
These studies demonstrate that femtoscopy provides a sensitive probe of the $N\Lambda_c$ interaction, allowing discrimination between different
interaction scenarios, including different strengths of attraction~\cite{Haidenbauer:2020kwo} and possible repulsive behavior~\cite{Zheng:2026qtk}.
In addition, explore the $N \Lambda_c$ interaction is also essential for investigating charmed hypernuclei~\cite{Dover:1977jw,Krein:2017usp,Hosaka:2016ypm,Gibson:1983zw}. 
Analogous to the hypertriton $^3_{\Lambda}\mathrm{H}$, whose structure is governed by the $\Lambda N$ interaction~\cite{Haidenbauer:2019boi,STAR:2022fnj,Chen:2023mel,ALICE:2022sco,Ma:2023}, the $N \Lambda_c$ interaction is expected to play a dominant role in determining the formation and properties of charmed hypernuclei~\cite{Vidana:2019amb,Haidenbauer:2020uci,Wu:2020nin,Liu:2023txn,Yang:2024ats,Bando:1981ti}.

In our previous work, the $N\Lambda_c$ interaction was investigated within the framework of the QDCSM~\cite{Huang:2013zva,Wu:2023yux}. 
By varying the color screening parameter $\mu$, different interaction strengths between the nucleon and $\Lambda_c$ were obtained~\cite{Huang:2013zva}.
The corresponding interaction potentials were further constructed by fitting the calculated scattering parameters with a Gaussian form. 
Based on these effective interactions, the possible existence of the $pn\Lambda_c$ bound states was investigated through a three-body calculation~\cite{Wu:2023yux}.
Within the same framework, we have investigated femtoscopic correlations of several hadronic systems~\cite{Yan:2024aap,Yan:2025hpa,Yan:2026yrd}, including $p\Omega$, $p\bar{\Omega}$, and $\Omega\phi$, demonstrating the applicability of this approach in exploring hadron-hadron interactions through correlation measurements.
In the present work, we extend this study by calculating the $p\Lambda_c$ femtoscopic correlation functions using these interaction potentials. 
Since the $p\Lambda_c$ system represents a two-body subsystem of the $pn\Lambda_c$ system, femtoscopic correlation measurements provide a unique opportunity to investigate the $N\Lambda_c$ interaction and to explore its connection with the possible $^3_{\Lambda_c}\mathrm{H}$ bound states.
By comparing the correlation functions corresponding to different interaction strengths, we investigate whether the possible formation of the heavy-flavor hypernucleus can be reflected in $p\Lambda_c$ femtoscopy.

This paper is organized as follows.
The theoretical framework for the $N\Lambda_c$ interaction and the calculation of the $p\Lambda_c$ femtoscopic correlation function are
introduced in Sec.~\ref{2}.
The numerical results and discussions are presented in Sec.~\ref{3}, where the sensitivity of the correlation functions to different $N\Lambda_c$ interactions and their implications for possible $pn\Lambda_c$ bound states are explored.
Finally, a summary is given in Sec.~\ref{summary}.

\section{THEORETICAL FORMALISM}
\label{2}
\subsection{Quark delocalization color screening model}

The details of the QDCSM employed in the present work can be found in Refs.~\cite{Wang:1992wi,Wu:1998wu,Pang:2001xx}.
Here, we present the brief introduction of the model.
The model Hamiltonian is given by
\begin{align}
	H =  \sum_{i=1}^6 \left(m_i+\frac{\boldsymbol{p}_{i}^{2}}{2m_i}\right) -T_{\mathrm{c} . \mathrm{m}} + \sum_{j>i=1}^6 V(\boldsymbol{r}_{ij}),
\label{H}
\end{align}
where $m_i$ is the quark mass, $\boldsymbol{p}_{i}$ is the momentum of the quark, and $T_{\mathrm{c.m.}}$ is the center-of-mass kinetic energy.
The dynamics of the hexaquark system is driven by two-body potentials, including color confinement ($V_{\mathrm{CON}}$), perturbative one-gluon-exchange interaction ($V_{\mathrm{OGE}}$), and dynamical chiral symmetry breaking ($V_{\chi}$).
\begin{align}
	V(\boldsymbol{r}_{ij})=  V_{\mathrm{CON}}(\boldsymbol{r}_{ij})+V_{\mathrm{OGE}}(\boldsymbol{r}_{ij})+V_{\chi}(\boldsymbol{r}_{ij}).
\end{align}

Here, a phenomenological color screening confinement potential ($V_{\mathrm{CON}}$) is used as
\begin{align}
	V_{\mathrm{CON}}(\boldsymbol{r}_{ij}) = & -a_{c}\boldsymbol{\lambda}_{i}^{c} \cdot \boldsymbol{\lambda}_{j}^{c}\left[  f(\boldsymbol{r}_{ij})+V_{0}\right],
\label{CON}
\end{align}
\begin{align}
	f(\boldsymbol{r}_{ij}) = \left\{\begin{array}{l}
		\boldsymbol{r}_{i j}^{2}, ~~~~~~~~i,j ~\text {in the same baryon orbit} \\
		\frac{1-\text{e}^{-\mu \boldsymbol{r}_{i j}^{2}}}{\mu},  ~i,j ~\text{in different baryon orbits}
	\end{array}\right.   \nonumber
\end{align}
where $a_c$ and $V_{0}$ are model parameters, and $\boldsymbol{\lambda}^{c}$ stands for the SU(3) color Gell-Mann matrices.
The color screening parameter $\mu$ adopted in this work is varied over three values, 0.8, 1.0, and 1.2~fm$^{-2}$.
The one-gluon-exchange potential ($V_{\mathrm{OGE}}$) is written as
\begin{align}
	V_{\mathrm{OGE}}(\boldsymbol{r}_{ij})= &\, \frac{1}{4}\alpha_{s_{q_i q_j}} \boldsymbol{\lambda}_{i}^{c} \cdot \boldsymbol{\lambda}_{j}^{c}
	\left[\frac{1}{r_{i j}}-\frac{3}{4 m_i m_j r_{i j}^3} S_{i j}\right. \nonumber  \\
	& - \left. \frac{\pi}{2} \delta\left(\boldsymbol{r}_{i j}\right)\left(\frac{1}{m_{i}^{2}}+\frac{1}{m_{j}^{2}}+\frac{4 \boldsymbol{\sigma}_{i} \cdot \boldsymbol{\sigma}_{j}}{3 m_{i} m_{j}}\right)\right], \\
	S_{i j}=&\, \frac{\left(\boldsymbol{\sigma}_i \cdot \boldsymbol{r}_{i j}\right)\left(\boldsymbol{\sigma}_j \cdot \boldsymbol{r}_{i j}\right)}{r_{i j}^2}-\frac{1}{3} \boldsymbol{\sigma}_i \cdot \boldsymbol{\sigma}_j
\label{OGE}
\end{align}
where $\alpha_{s}$ is the quark-gluon coupling constant, $\boldsymbol{\sigma}$ is the Pauli matrices, and $S_{i j}$ is the tensor operator.

The dynamical breaking of chiral symmetry results in the SU(3) Goldstone boson exchange interactions appear between constituent light quarks $u, d$, and $s$.
Hence, the chiral interaction is expressed as
\begin{align}
   V_{\chi}(\boldsymbol{r}_{ij})= &\,  V^{\pi}\left(\boldsymbol{r}_{i j}\right) \sum_{a=1}^{3} \boldsymbol{\lambda}_{i}^{a} \cdot \boldsymbol{\lambda}_{j}^{a}+V^{K}\left(\boldsymbol{r}_{i j}\right) \sum_{a=4}^{7} \boldsymbol{\lambda}_{i}^{a} \cdot \boldsymbol{\lambda}_{j}^{a} \nonumber \\
   &+V^{\eta}\left(\boldsymbol{r}_{i j}\right)\left[\left(\boldsymbol{\lambda}_{i}^{8} \cdot \boldsymbol{\lambda}_{j}^{8}\right) \cos \theta_{p}-\left(\boldsymbol{\lambda}_{i}^{0} \cdot \boldsymbol{\lambda}_{j}^{0}\right) \sin \theta_{p}\right], 
\end{align}
The potentials $V^{\pi}(\boldsymbol{r}_{ij})$, $V^{K}(\boldsymbol{r}_{ij})$, and $V^{\eta}(\boldsymbol{r}_{ij})$ for the corresponding exchanges share a common generic form, which is given below for $\chi = \pi, K, \eta$
 \begin{align}
   V^\chi(\boldsymbol{r}_{i j}) = &\, \frac{1}{3} \alpha_\text{ch} \frac{\Lambda^2}{\Lambda^2-m_\chi^2} m_\chi   \nonumber \\
	  & \times  \left\{  \left[Y\left(m_\chi r_{i j}\right)-\frac{\Lambda^3}{m_\chi^3} Y\left(\Lambda r_{i j}\right)\right] \boldsymbol{\sigma}_i \cdot \boldsymbol{\sigma}_j  \right. \nonumber \\
	&\left.+\left[H\left(m_\chi r_{i j}\right)-\frac{\Lambda^3}{m_\chi^3} H\left(\Lambda r_{i j}\right)\right] S_{i j}\right\} \boldsymbol{F}_i \cdot \boldsymbol{F}_j,  
\end{align}
where $H(x)=\left(1+3 / x+3 / x^{2}\right) Y(x)$ and $Y(x) = e^{-x}/x$ are the standard Yukawa functions.
The physical $\eta$ meson is considered by introducing the angle $\theta_{p}$ instead of the octet one.
The $\boldsymbol{\lambda}^{a}$ denote the SU(3) flavor Gell-Mann matrices, and the flavor generators are defined as $\boldsymbol{F}=\boldsymbol{\lambda}^{a}/2$.
The values of $m_\pi$, $m_k$ and $m_\eta$ are the masses of the SU(3) Goldstone bosons, which adopt the experimental values~\cite{ParticleDataGroup:2026aaa}.

In addition, quark delocalization was introduced to enlarge the model variational space to take into account the mutual distortion or the internal excitations of nucleons in the course of interaction.
It is realized by specifying the single-particle orbital wave function of the QDCSM as a linear combination of left and right Gaussians, the single-particle orbital wave functions used in the ordinary quark cluster model
\begin{eqnarray}
	\psi_{\alpha}(\boldsymbol {S_{i}} ,\epsilon) & = & \left(
	\phi_{\alpha}(\boldsymbol {S_{i}})
	+ \epsilon \phi_{\alpha}(-\boldsymbol {S_{i}})\right) /N(\epsilon), \nonumber \\
	\psi_{\beta}(-\boldsymbol {S_{i}} ,\epsilon) & = &
	\left(\phi_{\beta}(-\boldsymbol {S_{i}})
	+ \epsilon \phi_{\beta}(\boldsymbol {S_{i}})\right) /N(\epsilon), \nonumber \\
	N(S_{i},\epsilon) & = & \sqrt{1+\epsilon^2+2\epsilon e^{-S_i^2/4b^2}}. \label{1q}
\end{eqnarray}
It is worth noting that the mixing parameter $\epsilon$ is not an adjusted one but determined variationally by the dynamics of the multiquark system itself.
In this way, the multiquark system chooses its favorable configuration in the interacting process.
This mechanism has been used to explain the crossover transition between the hadron phase and quark-gluon plasma phase~\cite{Xu:2007oam}.
For more details about the QDCSM framework and the complete model parameters, we refer the readers to Ref.~\cite{Huang:2013zva}. 
In the present work, the baryon wave functions are constructed based on the group theory, whose details can be found in Ref.~\cite{Yan:2024usf}.

\subsection{Two-particle correlation function}
\label{22}

Experimentally, the correlation function $C(\boldsymbol{k})$ can be measured based on:
\begin{align}
	C(\boldsymbol{k}) & = \xi(\boldsymbol{k}) \frac{N_{\text{same}}(\boldsymbol{k})}{N_{\text{mixed}}(\boldsymbol{k})},
\end{align}
where $N_{\text{same}}(\boldsymbol{k})$ and $N_{\text{mixed}}(\boldsymbol{k})$ represent the $\boldsymbol{k}$ distributions of hadron-hadron pairs produced in the same
and in different collisions, respectively, and $\xi(\boldsymbol{k})$ denotes the corrections for experimental effects.
In theoretical studies, the correlation function can be calculated using the Koonin--Pratt (KP) formula~\cite{Koonin:1977fh,Pratt:1990zq,Bauer:1992ffu}:
\begin{align}
	C(\boldsymbol{k}) & = \frac{N_{12}\left(\boldsymbol{p}_{1}, \boldsymbol{p}_{2}\right)}{N_{1}\left(\boldsymbol{p}_{1}\right) N_{2}\left(\boldsymbol{p}_{2}\right)} \nonumber \\
	& \simeq \frac{\int \mathrm{d}^{4} x_{1} \mathrm{~d}^{4} x_{2} S_{1}\left(x_{1}, \boldsymbol{p}_{1}\right) S_{2}\left(x_{2}, \boldsymbol{p}_{2}\right)|\Psi(\boldsymbol{r}, \boldsymbol{k})|^{2}}{\int \mathrm{d}^{4} x_{1} \mathrm{~d}^{4} x_{2} S_{1}\left(x_{1}, \boldsymbol{p}_{1}\right) S_{2}\left(x_{2}, \boldsymbol{p}_{2}\right)} \nonumber \\
	& \simeq \int \mathrm{d} \boldsymbol{r} S_{12}(r)|\Psi(\boldsymbol{r}, \boldsymbol{k})|^{2},
\label{ignore}
\end{align}
where $S_{i}(x_{i}, \boldsymbol{p}_{i})~(i = 1, 2)$ is the single particle source function of the hadron $i$ with momentum $\boldsymbol{p}_{i}$, $\boldsymbol{k} = (m_2 \boldsymbol{p}_{1} - m_1 \boldsymbol{p}_{2})/(m_1 + m_2)$ is the relative momentum in the center-of-mass of the pair $(\boldsymbol{p}_{1} + \boldsymbol{p}_{2} = 0)$, $\boldsymbol{r}$ is the relative coordinate with time difference correction, and $\Psi(\boldsymbol{r}, \boldsymbol{k})$ is the relative wave function in the two-body outgoing state with an asymptotic relative momentum $\boldsymbol{k}$.
In the case where we can ignore the time difference of the emission and the momentum dependence of the source, we integrate out the center-of-mass coordinate and obtain Eq.~(\ref{ignore}), where $S_{12}(r)$ is the normalized pair source function in the relative coordinate, given by the expression:
\begin{align}
	S_{12}(r) = \frac{1}{(4 \pi R^2)^{3/2}} \text{exp}(-\frac{r^2}{4R^2}),
\label{source}
\end{align}
where $R$ is the size parameter of the source.
Thus, two important factors of the correlation function are included in Eq.~(\ref{ignore}): the collision system, which is related to the source function $S_{12}(r)$, and the two-particle interaction, which is embedded in the relative wave function $\Psi(\boldsymbol{r}, \boldsymbol{k})$.

For a pair of non-identical particles, such as proton--$\Lambda_c$, the correlation function at low relative momentum is dominated by the $S$-wave contribution. 
Therefore, only the modification of the $S$-wave component induced by the strong interaction is considered, and the relative wave function can be expressed as
\begin{align}
	\Psi_{p \Lambda_c}(\boldsymbol{r},\boldsymbol{k}) = \psi^C(\boldsymbol{r},\boldsymbol{k}) -\psi^C_0(r,k) +\psi_{p\Lambda_c}(r,k),
\end{align}
where $\psi^C(\boldsymbol{r},\boldsymbol{k})$ represents the Coulomb wave function, and $\psi^C_0(r,k)$ denotes its $S$-wave component. 
The last term, $\psi_{p\Lambda_c}(r,k)$, represents the Coulomb-distorted $S$-wave scattering wave function including both the strong and Coulomb interactions.

Substituting the above wave function into the KP formula, the correlation function can be written as
\begin{align}
	C_{p\Lambda_c}(k) =&\, \int_0^\infty 4\pi r^2 \mathrm{d}r\, S_{12}(r) \int\frac{\mathrm{d}\Omega}{4\pi} |\psi^C(\boldsymbol r,\boldsymbol k)|^2 \nonumber\\
	&+  \int_0^\infty 4\pi r^2\mathrm{d}r\, S_{12}(r) \left[ |\psi_{p\Lambda_c}(r,k)|^2 - |\psi^C_0(r,k)|^2 \right].
	\label{Ck}
\end{align}
Here, $\int \mathrm{d}\Omega$ denotes the angular integration over the relative direction between $\boldsymbol{k}$ and $\boldsymbol{r}$. 
The first term represents the contribution from the Coulomb interaction, while the second term accounts for the modification of the Coulomb $S$-wave component due to  the strong interaction.

The scattering wave function $\Psi_{p\Lambda_c}(\boldsymbol r,\boldsymbol k)$ can be obtained by solving the Schr\"odinger equation. A similar approach, where the correlation function is calculated from the scattering wave functions obtained by solving the Schr\"odinger equation, has been implemented in the femtoscopic correlation analysis tool (CATS)~\cite{Mihaylov:2018rva},
\begin{align}
	-\frac{\hbar^2}{2\mu}\nabla^2 \Psi_{p\Lambda_c}(\boldsymbol r,\boldsymbol k) + V(r)\Psi_{p\Lambda_c}(\boldsymbol r,\boldsymbol k) = E\Psi_{p\Lambda_c}(\boldsymbol r,\boldsymbol k),
\end{align}
where $\mu=m_pm_{\Lambda_c}/(m_p+m_{\Lambda_c})$ is the reduced mass of the $p\Lambda_c$ system.
The scattering wave function can generally be expanded into partial waves,
\begin{align}
	\Psi(\boldsymbol{r},\boldsymbol{k}) = \sum_{l,m} R_l(r,k)Y_l^m(\hat{\boldsymbol{r}}).
\end{align}
Since the low-momentum correlation function is dominated by the $S$-wave contribution, only the $l=0$ component is retained in the present work. The corresponding radial Schrödinger equation is given by
\begin{align}
	-\frac{\hbar^2}{2\mu} \frac{\mathrm{d}^2u_k(r)}{\mathrm{d}r^2} + V(r)u_k(r) = Eu_k(r),
	\label{eq}
\end{align}
where $E=\hbar^2k^2/(2\mu)$ and $u_k(r)=rR_k(r)$.

Using the obtained scattering wave functions, the correlation function $C_{p\Lambda_c}(k)$ for a given spin channel can be calculated through Eq.~(\ref{Ck}). 
This coordinate-space method, based on solving the Schr\"odinger equation, has been widely applied in femtoscopic correlation studies of various hadronic systems~\cite{Morita:2014kza,Ohnishi:2021ger}. 
Alternatively, the scattering wave functions can also be obtained by solving the Lippmann--Schwinger or Bethe--Salpeter equation in momentum space~\cite{Haidenbauer:2018jvl,Liu:2023uly}.

Since the experimentally measured correlation function is spin averaged, the theoretical correlation function should be obtained by averaging over different spin channels. The weights are determined by the spin multiplicities $(2J+1)$:
\begin{align}
	C_{p\Lambda_c}(k) = \frac{1}{4}C_{p\Lambda_c}^{J=0}(k) + \frac{3}{4}C_{p\Lambda_c}^{J=1}(k).
	\label{average}
\end{align}

\section{RESULTS AND DISCUSSION}
\label{3}

In the previous study, the $N\Lambda_c$ interaction was investigated within the framework of the QDCSM~\cite{Huang:2013zva,Wu:2023yux}. 
The resonating group method (RGM) ~\cite{Wheeler:1937zza,Yan:2023tvl} was employed to derive the interaction between the nucleon and $\Lambda_c$ baryon, where the coupled channels listed in Table~\ref{channels} were included.
Three sets of the color screening parameter $\mu$, i.e., $\mu=0.8$, $1.0$, and $1.2$ fm$^{-2}$, were considered to generate different interaction strengths. 
The energy spectra were analyzed by solving the coupled-channel RGM equation. 
It was found that, after including the $D$-wave channel couplings, the lowest energy of the $N\Lambda_c$ system remains above the $N\Lambda_c$ threshold and approaches the threshold with increasing model calculating space. 
This indicates that no $N\Lambda_c$ bound state is formed within the considered model~\cite{Huang:2013zva}.

\begin{table}[htb]
	\caption{Coupled channels included in the $N\Lambda_c$ interaction.}
	\begin{tabular}{l c}
		\hline \hline
		$J^P$ & Coupled channels  \\ \hline
		$0^+$ & $N \Lambda_c\left({ }^1 S_0\right), N \Sigma_c\left({ }^1 S_0\right), N \Sigma_c^*\left({ }^5 D_0\right)$   \\
		$1^+$ & $N \Lambda_c\left({ }^3 S_1\right), N \Sigma_c\left({ }^3 S_1\right), N \Sigma_c^*\left({ }^3 S_1\right)$,  \\ 
		~&$N \Lambda_c\left({ }^3 D_1\right), N \Sigma_c\left({ }^3 D_1\right), N \Sigma_c^*\left({ }^3 D_1\right), N \Sigma_c^*\left({ }^5 D_1\right)$ \\
		\hline \hline
		\label{channels}
	\end{tabular}
\end{table}

To further characterize the $N\Lambda_c$ interaction, the scattering phase shifts were calculated using the Kohn-Hulth\'en-Kato (KHK) method~\cite{Kamimura:1977okl}, from which the scattering length $a_0$ and effective range $r_{\text{eff}}$ were extracted through the low-energy expansion
\begin{align}
	k \cot \delta  =  -\frac{1}{a_{0}}+\frac{1}{2}r_{\text{eff}}k^{2}+{\cal O}(k^{4}),
\end{align}
where $k$ is the momentum of the relative motion with $k=\sqrt{2\mu E_{\mbox{c.m.}}}$, $\mu$ is the reduced mass of two baryons, and $E_{\mbox{c.m.}}$ is the incident energy; $\delta$ is the low-energy scattering phase shifts.
The corresponding scattering lengths and effective ranges for the $J^P=0^+$ and $1^+$ channels are summarized in Table~\ref{length}. 
These scattering parameters provide important information on the near-threshold behavior of the $N\Lambda_c$ interaction.
\begin{table}[htb]
	\caption{The scattering length $a_0$ and effective range $r_{\mathrm{eff}}$ of the $J^P=0^+$ and $1^+$ $N\Lambda_c$ systems. The color screening parameter $\mu$ is given in fm$^{-2}$.}
	\begin{tabular}{l c c c}
		\hline \hline
		$\mu$ &~~~~$J^{P}$~~~~  & ~~~~$a_{0}$ (fm)~~~~ & $r_{\text{eff}}$ (fm)  \\ \hline
		\multirow{2}{*}{0.8} & $0^+$ & $-$1.57 & 5.06  \\
		~                    & $1^+$ & $-$1.92 & 3.70  \\ \hline
		\multirow{2}{*}{1.0} & $0^+$ & $-$4.88 & 3.62  \\
        ~                    & $1^+$ & $-$7.11 & 3.15  \\ \hline
		\multirow{2}{*}{1.2} & $0^+$ & $-$2.58 & 3.65 \\
        ~                    & $1^+$ & $-$2.88 & 3.50  \\
		\hline \hline
		\label{length}
	\end{tabular}
\end{table}
Based on the obtained scattering parameters, the $N\Lambda_c$ interaction was parameterized by a two-range Gaussian potential,
\begin{align}
	V_{\text{eff}}(r) = V_1\exp(-r^2/b_1^2) + V_2\exp(-r^2/b_2^2),
\end{align}
where the fitted parameters are listed in Table~\ref{parameter}.
It should be noted that the above calculations focus on the strong interaction between the two hadrons, and therefore the Coulomb interaction is not included.

\begin{table}
\caption{Parameters of the effective $N\Lambda_c$ potential $V_{\rm eff}(r)$ defined in Eq.~(8), with $\mu$ in fm$^{-2}$, $V_1, V_2$ in MeV, and $b_1, b_2$ in fm, taken from Ref.~\cite{Wu:2023yux}.}
\begin{tabular}{c c c c c c c}
\hline \hline
 ~ & \multicolumn{3}{c}{$J^P=0^+$}  & \multicolumn{3}{c}{$J^P=1^+$} \\ \hline
    $\mu$  & ~~~0.8~~~ & ~~~1.0~~~ & ~~~1.2~~~ & ~~~0.8~~~ & ~~~1.0~~~ & ~~~1.2~~~ \\
    $V_1$  & $-$115.5 & $-$42.0 & $-$223.5 & $-$255.0 & $-$216.4 & $-$226.5  \\
    $b_1$  & 1.29 & 1.58 & 1.20 & 1.12 & 1.30 & 1.20  \\
    $V_2$  & 160.6 & 81.3 & 261.2 & 292.8 & 253.3 & 262.6  \\
    $b_2$  & 1.05 & 0.91 & 1.05 & 1.01 & 1.12 & 1.05 \\
\hline \hline
\label{parameter}
\end{tabular}
\end{table}

These effective $N\Lambda_c$ interactions were previously used to investigate the possible existence of charmed hypernuclear bound states. 
In Ref.~\cite{Wu:2023yux}, the $NN\Lambda_c$ three-body Schr\"odinger equation was solved within the Gaussian expansion method (GEM)~\cite{Hiyama:2003cu,Hiyama:2019kpw}, where the wave function was expanded as
\begin{align}
	\Psi_{JM} (NN\Lambda_c)= & \sum_{c=1}^{2} \sum_{s, S, L} \sum_{n_{1}, l_{1}} \sum_{n_{2}, l_{2}} C_{\gamma}^{(c)} \mathcal{A} \nonumber \\ 
	& \times\left\{\left\{\left(\phi_{n_{1} l_{1}}^{(c)}\left(\boldsymbol{r}_{c}\right) \psi_{n_{2} l_{2}}^{(c)}\left(\boldsymbol{R}_{c}\right)\right)_{L}\right.\right. \nonumber \\
	& \left.\left.\times\left[\left(\chi_{1 / 2}^{1} \chi_{1 / 2}^{2}\right)_{s} \chi_{1 / 2}^{3}\right]_{S}\right\}_{J M}\left(\tau_{1 / 2}^{1} \tau_{1 / 2}^{2}\right)_{T T_{z}}\right\},
\end{align}
where $c$ labels the two sets of Jacobi coordinates, $\mathcal{A}$ is the antisymmetrization operator for the two nucleons, and $\gamma$ denotes the set of quantum numbers $\{L,s,S,n_1,l_1,n_2,l_2\}$. Here, $\chi$ and $\tau$ represent the spin and isospin wave functions, respectively. 
The spatial wave functions were expanded in Gaussian basis functions,
\begin{align}
	\phi_{n_1 l_1 m_1}(\boldsymbol{r}) &= r^{l_1}  \text{e}^{-(r/r_{n_1})^2}   Y_{l_1 m_1}(\hat{\boldsymbol{r}}), \nonumber \\
	\psi_{n_2 l_2 m_2}(\boldsymbol{R}) &= R^{l_2}  \text{e}^{-(R/R_{n_2})^2}   Y_{l_2 m_2}(\hat{\boldsymbol{R}}),
\end{align}
where the Gaussian ranges $r_{n_1}$ and $R_{n_2}$ were chosen in geometric progression.

In the calculation of Ref.~\cite{Wu:2023yux}, the Coulomb interaction between the charged particles was also included. 
The calculated $\Lambda_c$ separation energies $B_{\Lambda_c}$ of the $^3_{\Lambda_c}\mathrm{H}$ system are summarized in Table~\ref{separation}. 
The results reveal three different binding scenarios for the $^3_{\Lambda_c}\mathrm{H}$ system. 
For $\mu=0.8~\mathrm{fm}^{-2}$, no bound state is obtained in either the $J^P=1/2^+$ or $3/2^+$ channel. 
The $\mu=1.2~\mathrm{fm}^{-2}$ case produces shallow bound states with $\Lambda_c$ separation energies of $0.08$ MeV and $0.16$ MeV for the $J^P=1/2^+$ and $3/2^+$ states,
respectively. 
In contrast, the $\mu=1.0~\mathrm{fm}^{-2}$ case leads to more deeply bound states, with separation energies of $0.85$ MeV and $1.31$ MeV for the $J^P=1/2^+$ and $3/2^+$ states, respectively. More details of the calculation can be found in Ref.~\cite{Wu:2023yux}.

\begin{table}
	\caption{Separation energies of the $^3_{\Lambda_c}\mathrm{H}$ system for different values of the color screening parameter $\mu$, where
		$B_{\Lambda_c}=M_d + M_{\Lambda_c} - M_{pn\Lambda_c}$ denotes the $\Lambda_c$ separation energy. 
		The energies are given in MeV, and $\mu$ is given in fm$^{-2}$.}
	\begin{tabular}{l c c c}
		\hline \hline
		& ~~$\mu$ = 0.8 ~~  & ~~$\mu$ = 1.0 ~~  & ~~$\mu$ = 1.2 ~~   \\ \hline
		$B_{\Lambda_c}$ ($1/2^+$)  & UB & 0.85 & 0.08   \\
		$B_{\Lambda_c}$ ($3/2^+$)  & UB & 1.31 & 0.16  \\ 
		\hline \hline  
		\label{separation}
	\end{tabular}
\end{table}

Based on the effective $N\Lambda_c$ interactions obtained in the previous study, we further investigate the corresponding $p\Lambda_c$ femtoscopic correlations in the present work.
By solving the two-body Schr\"odinger equation with these interactions, the scattering wave functions are obtained and used in the KP formula to calculate the correlation functions.
This provides a direct connection between the theoretically predicted $^3_{\Lambda_c}\mathrm{H}$ binding properties and experimentally measurable $p\Lambda_c$ femtoscopic observables.

\begin{figure}[htb]
	\centering
	\includegraphics[width=9.5cm]{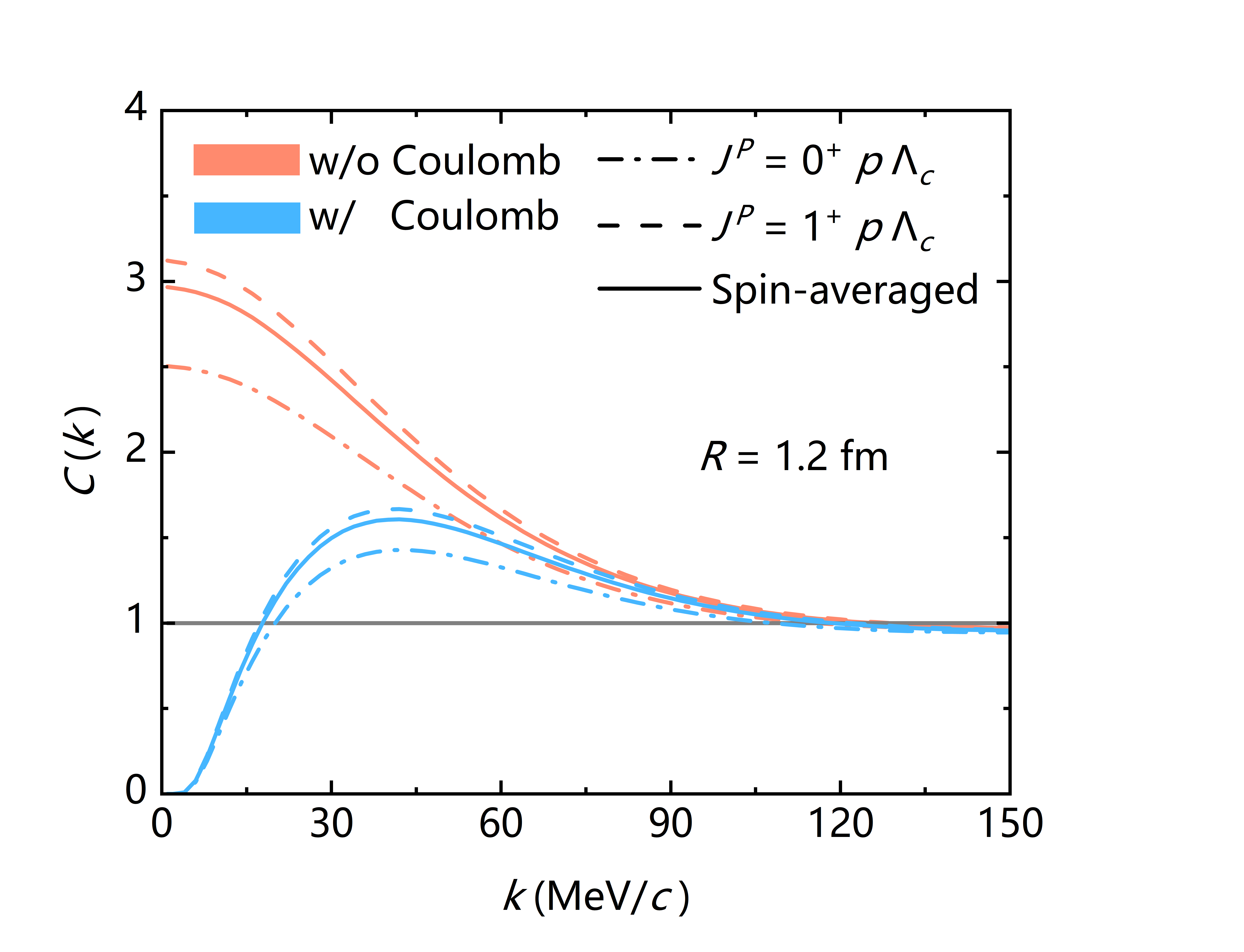}\
	\caption{$p\Lambda_c$ correlation functions in the $J^P=0^+$ and $J^P=1^+$ channels with and without the Coulomb interaction for the $\mu=0.8$ fm$^{-2}$ scenario. The spin-averaged correlation function is also shown. The source size is fixed at $R=1.2$ fm.}
	\label{coulomb}
\end{figure}

We first investigate the impact of the $p\Lambda_c$ interaction on the femtoscopic correlation function. 
Since the $N\Lambda_c$ interaction obtained in the previous study contains two spin channels, the correlation functions for the
$J^P=0^+$ and $J^P=1^+$ states are calculated separately.
As an example, the interaction corresponding to $\mu=0.8$ fm$^{-2}$ is selected, which does not support a $^3_{\Lambda_c}\mathrm{H}$ bound state.
The results with and without the Coulomb interaction for a source size of $R=1.2$ fm are shown in Fig.~\ref{coulomb}.

Before considering the Coulomb interaction, it can be seen that both spin channels exhibit an enhancement of the correlation function above unity at low relative momenta, indicating that the strong $p\Lambda_c$ interaction is attractive in both the $J^P=0^+$ and $1^+$ channels.
Moreover, the enhancement in the $J^P=1^+$ channel is stronger than that in the $J^P=0^+$ channel, suggesting a stronger attractive interaction
in the triplet channel.
This behavior is consistent with the scattering parameters listed in Table~\ref{length}, where the $J^P=1^+$ channel has a larger magnitude
of the scattering length compared with the $J^P=0^+$ channel.
Furthermore, a similar hierarchy between the two spin channels was also observed in Ref.~\cite{Maeda:2015hxa}, where the
$J^P=1^+$ interaction was found to be stronger than the $J^P=0^+$ case.
When the Coulomb interaction is included, the correlation functions are significantly suppressed in the low-$k$ region.
This suppression originates from the repulsive long-range Coulomb interaction between the positively charged proton and $\Lambda_c^+$.

\begin{figure}[htb]
	\centering
	\includegraphics[width=9.5cm]{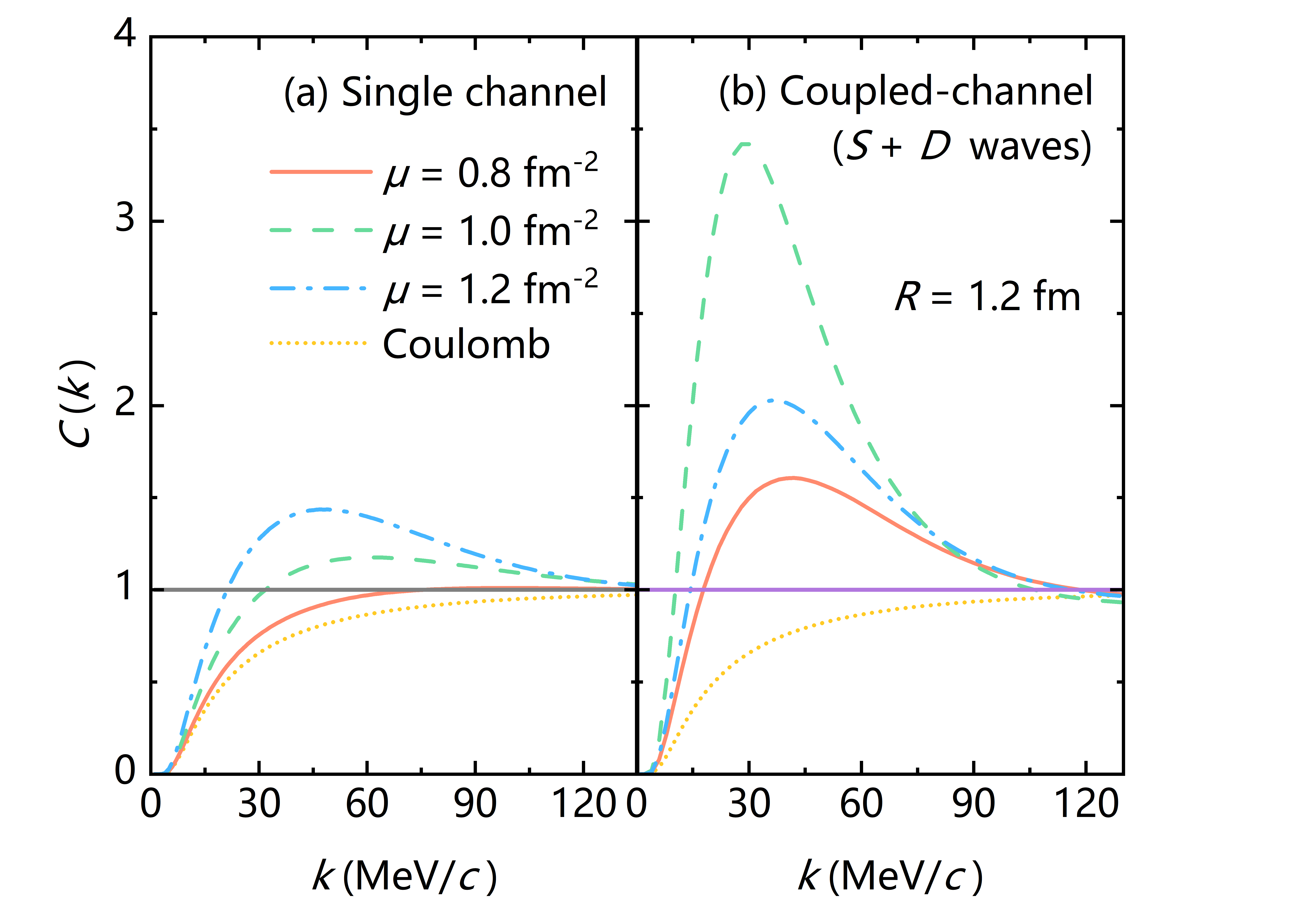}\
	\caption{$p\Lambda_c$ correlation functions corresponding to different $^3_{\Lambda_c}\mathrm{H}$ binding scenarios characterized by the color screening parameter $\mu$. Results from the single-channel and coupled-channel calculations are shown in panels (a) and (b), respectively. The coupled-channel results include both $S$- and $D$-wave contributions, and the source size is fixed at $R=1.2$ fm.}
	\label{coupling}
\end{figure}

Furthermore, the effects of coupled channels and $D$-wave contributions are investigated in the femtoscopic correlation function.
In the coupled-channel calculation, the effects of the $N\Sigma_c$, $N\Sigma_c^*$ channels and the $D$-wave components are included in the
low-energy $N\Lambda_c$ scattering parameters.
By constructing an effective $S$-wave $N\Lambda_c$ interaction that reproduces the scattering length and effective range obtained from the
full coupled-channel calculation, the coupled-channel and $D$-wave effects are incorporated into the scattering wave functions used for
the correlation function.
The corresponding correlation functions obtained from the single-channel and coupled-channel calculations are presented in Fig.~\ref{coupling}.

Compared with the single-channel results, the coupled-channel correlations exhibit a significantly stronger enhancement at low relative momentum for all three values of $\mu$.
This indicates that the coupling to the $N\Sigma_c$, $N\Sigma_c^*$ and $D$-wave channels provides additional attraction to the $N\Lambda_c$ interaction.
Since the $N\Lambda_c$ channel is the lowest threshold among the considered channels, the coupling effects with other channels can further lower the energy of the system and enhance the attractive behavior.

For the single-channel calculation, the attractive behavior becomes stronger with increasing $\mu$.
This behavior is consistent with the QDCSM framework, where the color screening parameter $\mu$ affects the strength of the screened confinement interaction.
A larger $\mu$ usually leads to a more attractive $N\Lambda_c$ interaction.
However, after including the coupled-channel effects, the dependence on $\mu$ becomes non-monotonic.
In particular, the strongest correlation enhancement appears for $\mu=1.0$ fm$^{-2}$ rather than $\mu=1.2$ fm$^{-2}$.
This demonstrates that the coupled-channel dynamics plays an essential role in determining the low-energy $N\Lambda_c$ interaction and cannot be simply understood from the strength of the diagonal interaction alone.

\begin{figure}[htb]
	\centering
	\includegraphics[width=9.5cm]{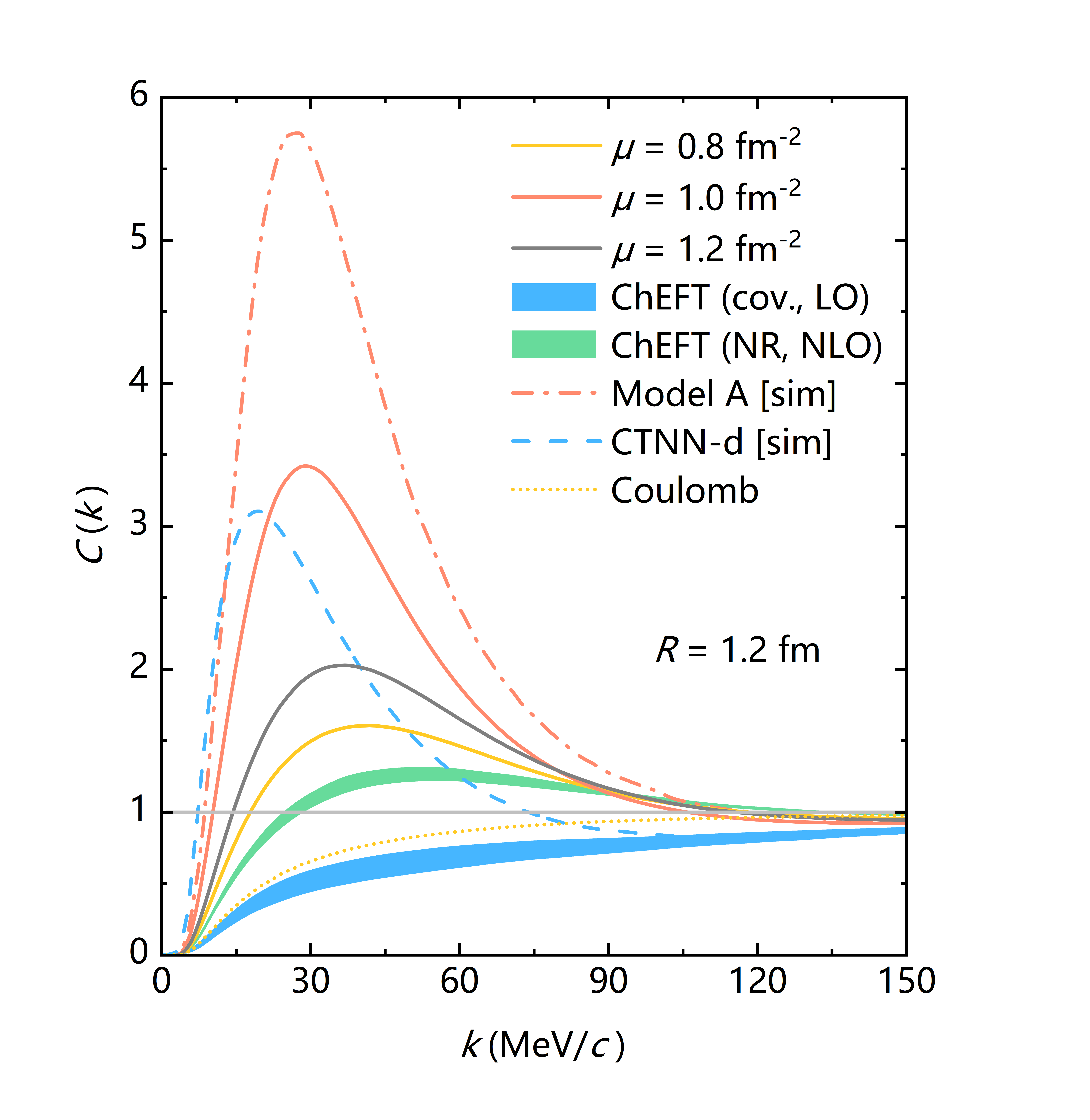}\
	\caption{Calculated $p\Lambda_c$ femtoscopic correlation functions. 
		The effective potentials correspond to: no bound $^3_{\Lambda_c}\mathrm{H}$ state for $\mu=0.8~\mathrm{fm}^{-2}$ (solid yellow line), shallow $^3_{\Lambda_c}\mathrm{H}$ bound states for $\mu=1.2~\mathrm{fm}^{-2}$ (solid gray line), and more deeply $^3_{\Lambda_c}\mathrm{H}$ bound states for $\mu=1.0~\mathrm{fm}^{-2}$ (solid red line).
		The results from other $p\Lambda_c$ interactions, including covariant ChEFT (blue band)~\cite{Zheng:2026qtk}, nonrelativistic ChEFT at next-to-leading order (green band)~\cite{Haidenbauer:2020kwo}, Model A (red dash-dotted line)~\cite{Haidenbauer:2020kwo,Vidana:2019amb}, and CTNN-d (blue dash line)~\cite{Maeda:2015hxa,Haidenbauer:2020kwo} are also presented.}
	\label{compare}
\end{figure}

After incorporating the Coulomb interaction and the coupled-channel and $S$-$D$ wave effects into the effective $p\Lambda_c$ interactions, together with the spin averaging, the $p\Lambda_c$ femtoscopic correlation functions can be calculated using the effective interactions  obtained from the QDCSM framework. 
Fig.~\ref{compare} presents the calculated correlation functions for the three interaction scenarios characterized by different color screening parameters $\mu=0.8$, $1.0$, and $1.2~\mathrm{fm}^{-2}$.
These three cases correspond to different possible binding properties of the $^3_{\Lambda_c}\mathrm{H}$ system, as discussed above.
For comparison, the results from other theoretical approaches, including covariant ChEFT~\cite{Zheng:2026qtk}, nonrelativistic ChEFT at next-to-leading order~\cite{Haidenbauer:2020kwo}, a SU(4) extended Jülich meson-exchange model (Model A)~\cite{Vidana:2019amb}, and a phenomenological baryon–nucleon potential model (CTNN-d)~\cite{Maeda:2015hxa}, are also displayed.
The latter two results are reproduced from Ref.~\cite{Haidenbauer:2020kwo}.

For $\mu=0.8~\mathrm{fm}^{-2}$, the corresponding $pn\Lambda_c$ system does not support a $^3_{\Lambda_c}\mathrm{H}$ bound state, while the $\mu=1.2~\mathrm{fm}^{-2}$ 
interaction produces shallow bound $^3_{\Lambda_c}\mathrm{H}$ states with a separation energy of approximately $0.1$ MeV.
The $\mu=1.0~\mathrm{fm}^{-2}$ case leads to more deeply bound  $^3_{\Lambda_c}\mathrm{H}$ states with a separation energy of about 
$1$ MeV.
Interestingly, these different binding scenarios result in clearly distinguishable behaviors in the $p\Lambda_c$ correlation functions.
The correlation enhancement at low relative momentum is strongly correlated with the strength of the attractive $p\Lambda_c$ interaction.
Among the three cases, the $\mu=1.0~\mathrm{fm}^{-2}$ interaction produces the strongest enhancement, followed by the $\mu=1.2~\mathrm{fm}^{-2}$ 
and $\mu=0.8~\mathrm{fm}^{-2}$ cases.
This behavior reflects the different near-threshold properties of the $p\Lambda_c$ interaction, where a stronger attraction and a larger scattering effect lead to a more pronounced modification of the two-particle wave function and consequently a stronger femtoscopic signal.

The $\mu=1.0~\mathrm{fm}^{-2}$ case exhibits a much stronger correlation enhancement than the $\mu=1.2~\mathrm{fm}^{-2}$ case, although both interactions support bound $^3_{\Lambda_c}\mathrm{H}$ states.
Specifically, the correlation function reaches a maximum value of approximately $C(k)\simeq 3.4$ for $\mu=1.0~\mathrm{fm}^{-2}$, while it is reduced to about $C(k)\simeq 2.0$ for the shallow-bound scenario with $\mu=1.2~\mathrm{fm}^{-2}$.
Meanwhile, even the $\mu=0.8~\mathrm{fm}^{-2}$ interaction, which does not support a bound $^3_{\Lambda_c}\mathrm{H}$ state, still produces a noticeable enhancement with a maximum value of $C(k)\simeq 1.6$ due to its attractive nature.
These results demonstrate that the femtoscopic observable is sensitive to the detailed low-energy behavior of the $p\Lambda_c$ interaction.

Furthermore, the comparison with other theoretical approaches provides a broader perspective on the $p\Lambda_c$ interaction and its corresponding femtoscopic signatures.
For the phenomenological potential model, Model A~\cite{Vidana:2019amb} predicts a much stronger attractive $p\Lambda_c$ interaction, with the scattering lengths of $a_0(^1S_0)=-2.60$ fm and $a_0(^3S_1)=-15.87$ fm.
Consequently, the correlation function obtained from this interaction exhibits a pronounced enhancement, reaching values above $C(k)=5.7$, as shown by the red dash-dotted curve in Fig.~\ref{compare}.
Although both Model A and our QDCSM interactions indicate attractive $p\Lambda_c$ interactions, they are not sufficiently attractive to form a two-body $p\Lambda_c$ bound state.
Therefore, their correlation functions do not show a significant suppression below the pure Coulomb case at higher relative momentum.

In contrast, the CTNN-d interaction predicts bound $\Lambda_c N$ states in both the $J^P=0^+$ and $1^+$ channels after including channel couplings.
When applied to the three-body $pn\Lambda_c$ system, this interaction also leads to deeply bound states with binding energies exceeding 20 MeV.
The corresponding correlation function, shown by the blue dashed curve in Fig.~\ref{compare}, exhibits a clear suppression below the Coulomb-only result at $k\sim100$ MeV/$c$.
Such a suppression is associated with the presence of bound states, which reduces the probability of observing scattering pairs at finite relative momentum.
Such a strong attractive interaction in the presence of the repulsive Coulomb interaction generally results in a correlation function with a strong sensitivity to the source size.

The correlation functions obtained from ChEFT interactions are also compared.
The nonrelativistic ChEFT interaction at next-to-leading order (NLO)~\cite{Haidenbauer:2020kwo}, shown by the green band in Fig.~\ref{compare}, predicts weakly attractive $p\Lambda_c$ interactions in both the spin-singlet and spin-triplet channels, leading to a moderate correlation enhancement.
For the covariant ChEFT interaction~\cite{Zheng:2026qtk}, their analysis shows that the interaction remains weakly attractive in the spin-singlet $^1S_0$ channel, while the spin-triplet $^3S_1$ channel becomes sensitive to coupled-channel effects.
In particular, the inclusion of $S$-$D$ wave mixing changes the triplet interaction from attractive to repulsive.
Since the spin-averaged correlation function is dominated by the triplet contribution, the resulting correlation function exhibits a suppression compared with the Coulomb-only case, as shown by the blue band in Fig.~\ref{compare}.

The above results demonstrate that femtoscopic correlation functions provide a sensitive probe of the low-energy $p\Lambda_c$ interaction.
For a fixed source size, the magnitude of the correlation enhancement and its deviation from the pure Coulomb result provide important information on the nature of the strong interaction.
In general, an attractive strong interaction leads to an enhancement of the correlation function above the Coulomb-only result at low relative momentum, while a repulsive interaction tends to suppress the correlation function relative to the Coulomb-only result.
Furthermore, once an attractive strong interaction has been established, a suppression of the correlation function below the Coulomb-only result in certain momentum regions may be associated with bound-state effects.
Therefore, the detailed behavior of the femtoscopic correlation function can provide constraints on the low-energy $p\Lambda_c$ interaction and offer valuable guidance for theoretical scenarios related to possible $^3_{\Lambda_c}\mathrm{H}$ bound states.
However, such behavior should not be regarded as a unique signature of $^3_{\Lambda_c}\mathrm{H}$ bound states, since the correlation function is a spin-averaged two-body observable.
A direct connection to the three-body bound state requires a consistent description of all relevant $N\Lambda_c$ channels and the three-body dynamics.

\begin{figure*}[htb]
	\centering
	\includegraphics[width=18.5cm]{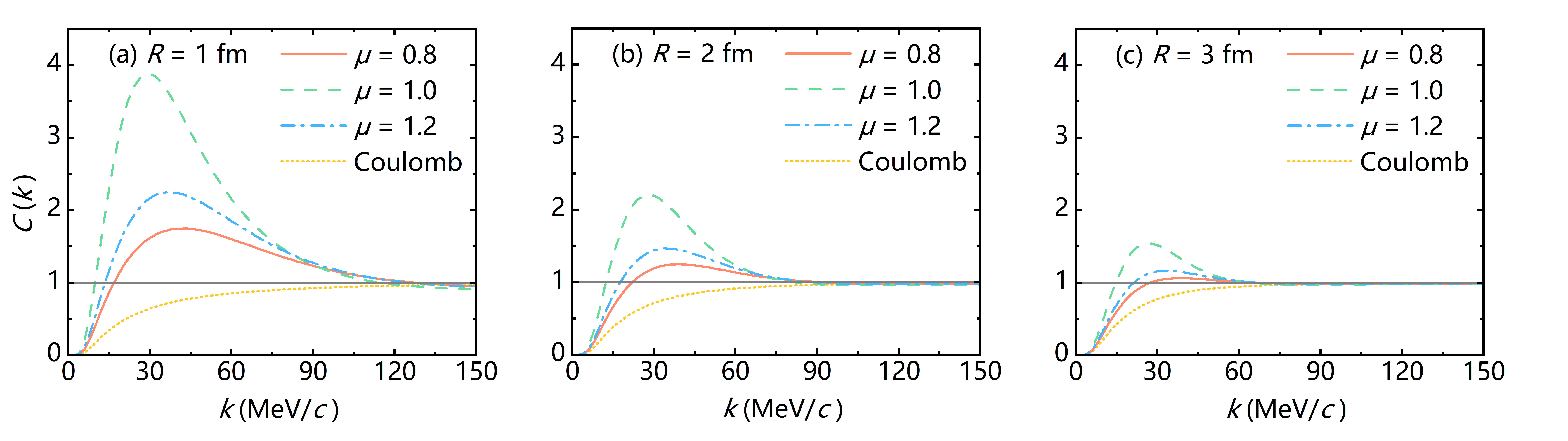}\
	\caption{Source-size dependence of the $p\Lambda_c$ femtoscopic correlation functions. The correlation functions calculated with different color screening parameters $\mu$, corresponding to different binding properties of the $^3_{\Lambda_c}\mathrm{H}$ system, are shown for source sizes of $R=1$, $2$, and $3$ fm. The Coulomb-only results are also presented.}
	\label{size} 
\end{figure*}

Since the femtoscopic correlation function is determined by the interplay between the source distribution and the final-state interaction, the source size plays a crucial role in controlling the sensitivity to the short-range interaction~\cite{ALICE:2023sjd,Mihaylov:2023pyl,VazquezDoce:2024nye,Wang:2025htd}.
The source-size dependence of the $p\Lambda_c$ correlation functions is further investigated in Fig.~\ref{size}, where the results for source radii of $R=1$, $2$, and $3$ fm are presented for the three interaction scenarios characterized by different color screening parameters.
As expected, the magnitude of the correlation signal decreases with increasing source size, since the influence of the short-range strong interaction becomes diluted for a larger emission source.
For relatively small source sizes, the three interaction scenarios considered in our QDCSM-based framework, namely the non-bound scenario ($\mu=0.8~\mathrm{fm}^{-2}$), the shallow-bound scenario ($\mu=1.2~\mathrm{fm}^{-2}$), and the more deeply bound scenario ($\mu=1.0~\mathrm{fm}^{-2}$), remain distinguishable through their different correlation behaviors.
This suggests that femtoscopic measurements with sufficiently small source sizes may provide additional constraints on the possible strength of the $p\Lambda_c$ interaction and offer useful guidance for exploring the existence of possible $^3_{\Lambda_c}\mathrm{H}$ bound states within specific theoretical frameworks.

\section{SUMMARY}
\label{summary}

In this work, we have investigated the $p\Lambda_c$ femtoscopic correlation function based on the $N\Lambda_c$ interactions obtained within the QDCSM framework.
By varying the color screening parameter $\mu$, three interaction scenarios with different strengths and corresponding binding properties of the $^3_{\Lambda_c}\mathrm{H}$ system are constructed.
The coupled-channel effects and $S$-$D$ wave mixing are incorporated through effective $S$-wave interactions that reproduce the low-energy scattering parameters.

Using these interactions together with the Coulomb interaction, the spin-averaged $p\Lambda_c$ correlation functions are calculated within the KP formalism.
The results show that the correlation function is highly sensitive to the low-energy behavior of the $p\Lambda_c$ interaction, and the coupled-channel dynamics can lead to significant modifications compared with the single-channel results.
The different interaction scenarios associated with unbound, shallow-bound, and more deeply bound $^3_{\Lambda_c}\mathrm{H}$ systems exhibit distinguishable correlation patterns within the present framework.

With the rapid development of femtoscopic measurements in relativistic heavy-ion collisions at the LHC and RHIC, such as those performed by the ALICE and STAR collaborations, correlation studies involving heavy-flavor hadrons are becoming an important tool for exploring hadronic interactions.
Although the $p\Lambda_c$ correlation function is a spin-averaged two-body observable and cannot uniquely determine the three-body bound state, it can provide valuable constraints on the low-energy $p\Lambda_c$ interaction and offer important guidance for future searches for possible heavy-flavor hypernuclei.

\acknowledgments{This work is supported partly by the National Natural Science Foundation of China under Contracts Nos. 12305087 and 12575088. 	
	Y.Y. is supported by the Scientific Research Foundation of Changzhou University of Information Technology under Grant No. SG210201B13003 and by the Changzhou Sci\&Tech Program under Grant No. 20260098.
	Q. W. is supported by the Natural Science Foundation of Jiangsu Province under grant No. BK20220122, the China Postdoctoral Science Foundation under grant No. 2024M751369, and the Jiangsu Funding Program for Excellent Postdoctoral Talent.}	

\setcounter{equation}{0}
\renewcommand\theequation{A\arabic{equation}}


\begin{thebibliography}{99}

% review	
\bibitem{Wiedemann:1999qn} U.~A.~Wiedemann and U.~W.~Heinz, Particle interferometry for relativistic heavy ion collisions, Phys. Rept. \textbf{319}, 145 (1999).	
\bibitem{Lisa:2005dd} M.~A.~Lisa, S.~Pratt, R.~Soltz and U.~Wiedemann, Femtoscopy in relativistic heavy ion collisions, Ann. Rev. Nucl. Part. Sci. \textbf{55}, 357 (2005).
\bibitem{ExHIC:2017smd} S.~Cho \textit{et al.} [ExHIC], Exotic hadrons from heavy ion collisions, Prog. Part. Nucl. Phys. \textbf{95}, 279 (2017).
\bibitem{Fabbietti:2020bfg} L.~Fabbietti, V.~Mantovani Sarti and O.~Vazquez Doce, Study of the Strong Interaction Among Hadrons with Correlations at the LHC, Ann. Rev. Nucl. Part. Sci. \textbf{71}, 377 (2021).
\bibitem{Liu:2024uxn} M.~Z.~Liu, Y.~W.~Pan, Z.~W.~Liu, T.~W.~Wu, J.~X.~Lu and L.~S.~Geng, Three ways to decipher the nature of exotic hadrons: Multiplets, three-body hadronic molecules, and correlation functions, Phys. Rept. \textbf{1108}, 1 (2025).
	
% NN
\bibitem{Koonin:1977fh} S.~E.~Koonin, Proton Pictures of High-Energy Nuclear Collisions, Phys. Lett. B \textbf{70}, 43 (1977).
\bibitem{Lednicky:1981su} R.~Lednicky and V.~L.~Lyuboshits, Final State Interaction Effect on Pairing Correlations Between Particles with Small Relative Momenta, Yad. Fiz. \textbf{35}, 1316 (1981).
\bibitem{STAR:2015kha} L.~Adamczyk \textit{et al.} [STAR], Measurement of Interaction between Antiprotons, Nature \textbf{527}, 345 (2015).
\bibitem{ALICE:2020ibs} S.~Acharya \textit{et al.} [ALICE], Search for a common baryon source in high-multiplicity pp collisions at the LHC, Phys. Lett. B \textbf{811}, 135849 (2020) [erratum: Phys. Lett. B \textbf{861}, 139233 (2025)].
\bibitem{ALICE:2025wuy} S.~Acharya \textit{et al.} [ALICE], Femtoscopic study of the proton-proton and proton-deuteron systems in heavy-ion collisions at the LHC, Phys. Lett. B \textbf{871}, 139921 (2025).
\bibitem{SRIT:2026qkv} Y.~J.~Wang \textit{et al.} [S$\pi$RIT], Large amplification of the isospin-dependence of proton emitting source size in radioactive heavy-ion collisions: a signal of $n$-$p$ correlation, [arXiv:2604.25107 [nucl-ex]].
\bibitem{Xi:2026vrp} B.~Xi, P.~Li, C.~Zhang, J.~Chen, S.~Y.~L.~T.~Zhang and Y.~G.~Ma, Proton-proton Femtoscopy as a Probe of Short-range Structure in High-Energy O+O Collisions, [arXiv:2608.01190 [nucl-th]].

% NY
\bibitem{ALICE:2019hdt} S.~Acharya \textit{et al.} [ALICE], First Observation of an Attractive Interaction between a Proton and a Cascade Baryon, Phys. Rev. Lett. \textbf{123}, 112002 (2019).
\bibitem{ALICE:2020mfd} A.~Collaboration \textit{et al.} [ALICE], Unveiling the strong interaction among hadrons at the LHC, Nature \textbf{588}, 232 (2020) [erratum: Nature \textbf{590}, E13 (2021)].
\bibitem{ALICE:2021njx} S.~Acharya \textit{et al.} [ALICE], Exploring the $N \Lambda$--$N \Sigma$ coupled system with high precision correlation techniques at the LHC, Phys. Lett. B \textbf{833}, 137272 (2022).
\bibitem{Fu:2024btw} B.~Fu [STAR], Measurement of $p$--$\Xi^-$ $(p$--$\bar{\Xi}^+)$ Correlation Function in Isobar and Au+Au Collisions at $\sqrt{s_{NN}} = 200$ GeV with the STAR Detector, EPJ Web Conf. \textbf{316}, 03010 (2025).
\bibitem{STAR:2005rpl} J.~Adams \textit{et al.} [STAR], Proton--$\Lambda$ correlations in central Au+Au collisions at $\sqrt{s_{NN}}$ = 200 GeV, Phys. Rev. C \textbf{74}, 064906 (2006).
\bibitem{STAR:2018uho} J.~Adam \textit{et al.} [STAR], The Proton--$\Omega$ correlation function in Au+Au collisions at $\sqrt{s_{NN}}$ = 200 GeV, Phys. Lett. B \textbf{790}, 490 (2019).
\bibitem{Ohnishi:2016elb} A.~Ohnishi, K.~Morita, K.~Miyahara and T.~Hyodo, Hadron--hadron correlation and interaction from heavy-ion collisions, Nucl. Phys. A \textbf{954}, 294 (2016).
\bibitem{Morita:2016auo} K.~Morita, A.~Ohnishi, F.~Etminan and T.~Hatsuda, Probing multistrange dibaryons with proton-omega correlations in high-energy heavy ion collisions, Phys. Rev. C \textbf{94}, 031901 (2016).
\bibitem{Hatsuda:2017uxk} T.~Hatsuda, K.~Morita, A.~Ohnishi and K.~Sasaki, $p\Xi^- $ Correlation in Relativistic Heavy Ion Collisions with Nucleon-Hyperon Interaction from Lattice QCD, Nucl. Phys. A \textbf{967}, 856 (2017).
\bibitem{Haidenbauer:2021zvr} J.~Haidenbauer and U.~G.~Mei{\ss}ner, Exploring the $\Sigma^+$p interaction by measurements of the correlation function, Phys. Lett. B \textbf{829}, 137074 (2022).
\bibitem{Garrido:2024pwi} E.~Garrido, A.~Kievsky, M.~Gattobigio, M.~Viviani, L.~E.~Marcucci, R.~Del Grande, L.~Fabbietti and D.~Melnichenko, $p \Lambda$ and $pp \Lambda$ correlation functions, Phys. Rev. C \textbf{110}, 054004 (2024).

% YY
\bibitem{ALICE:2018ysd} S.~Acharya \textit{et al.} [ALICE], $p$--$p$, $p$--$\Lambda$ and $\Lambda$--$\Lambda$ correlations studied via femtoscopy in pp reactions at $\sqrt{s}$ = 7 TeV, Phys. Rev. C \textbf{99}, 024001 (2019).
\bibitem{Ohnishi:1998at} A.~Ohnishi, Y.~Hirata, Y.~Nara, S.~Shinmura and Y.~Akaishi, Can we extract Lambda-Lambda interaction from two particle momentum correlation?, Nucl. Phys. A \textbf{670}, 297 (2000).
\bibitem{Morita:2014kza} K.~Morita, T.~Furumoto and A.~Ohnishi, $\Lambda\Lambda$ interaction from relativistic heavy-ion collisions, Phys. Rev. C \textbf{91}, 024916 (2015).
\bibitem{Haidenbauer:2018jvl} J.~Haidenbauer, Coupled-channel effects in hadron--hadron correlation functions, . Phys. A \textbf{981}, 1 (2019).
\bibitem{Morita:2019rph} K.~Morita, S.~Gongyo, T.~Hatsuda, T.~Hyodo, Y.~Kamiya and A.~Ohnishi, Probing $\Omega\Omega$ and $p\Omega$ dibaryons with femtoscopic correlations in relativistic heavy-ion collisions, Phys. Rev. C \textbf{101}, 015201 (2020).
\bibitem{Ohnishi:2021ger} A.~Ohnishi, Y.~Kamiya, K.~Sasaki, T.~Fukui, T.~Hatsuda, T.~Hyodo, K.~Morita and K.~Ogata, Femtoscopic Study of $N \Xi $ Interaction and Search for the H Dibaryon State Around the $N \Xi $ Threshold, Few Body Syst. \textbf{62}, 42 (2021).
\bibitem{Kamiya:2021hdb} Y.~Kamiya, K.~Sasaki, T.~Fukui, T.~Hyodo, K.~Morita, K.~Ogata, A.~Ohnishi and T.~Hatsuda, Femtoscopic study of coupled-channels $N \Xi$ and $\Lambda\Lambda$ interactions, Phys. Rev. C \textbf{105}, 014915 (2022).
\bibitem{Liu:2022nec} Z.~W.~Liu, K.~W.~Li and L.~S.~Geng, Strangeness S = $-$2 baryon-baryon interactions and femtoscopic correlation functions in covariant chiral effective field theory, Chin. Phys. C \textbf{47}, 024108 (2023).
\bibitem{STAR:2014dcy} L.~Adamczyk \textit{et al.} [STAR], $\Lambda\Lambda$ Correlation Function in Au+Au collisions at $\sqrt{s_{NN}}=$ 200 GeV, Phys. Rev. Lett. \textbf{114}, 022301 (2015).
\bibitem{ALICE:2022uso} S.~Acharya \textit{et al.} [ALICE], First measurement of the $\Lambda$--$\Xi$ interaction in proton--proton collisions at the LHC, Phys. Lett. B \textbf{844}, 137223 (2023).
\bibitem{Sarti:2025sdo} V.~M.~Sarti, Novel constraints on $\Lambda$--$\bar{\Lambda}$ and $p$--$\bar{\Lambda}$ interactions using correlation data, Eur. Phys. J. C \textbf{85}, 1068 (2025).

% charm
\bibitem{ALICE:2022enj} S.~Acharya \textit{et al.} [ALICE], First study of the two-body scattering involving charm hadrons, Phys. Rev. D \textbf{106}, 052010 (2022).
\bibitem{Kamiya:2022thy} Y.~Kamiya, T.~Hyodo and A.~Ohnishi, Femtoscopic study on $DD^*$ and $D\bar{D}^*$ interactions for $T_{cc}$ and X(3872), Eur. Phys. J. A \textbf{58}, 131 (2022).
\bibitem{Liu:2023uly} Z.~W.~Liu, J.~X.~Lu and L.~S.~Geng, Study of the DK interaction with femtoscopic correlation functions, Phys. Rev. D \textbf{107}, 074019 (2023).
\bibitem{Vidana:2023olz} I.~Vidana, A.~Feijoo, M.~Albaladejo, J.~Nieves and E.~Oset, Femtoscopic correlation function for the $T_{c}(3875)^+$ state, Phys. Lett. B \textbf{846}, 138201 (2023).
\bibitem{Albaladejo:2023pzq} M.~Albaladejo, J.~Nieves and E.~Ruiz-Arriola, Femtoscopic signatures of the lightest $S$-wave scalar open-charm mesons, Phys. Rev. D \textbf{108}, 014020 (2023).
\bibitem{Ikeno:2023ojl} N.~Ikeno, G.~Toledo and E.~Oset, Model independent analysis of femtoscopic correlation functions: An application to the $D_{s0}^*(2317)$, Phys. Lett. B \textbf{847}, 138281 (2023).
\bibitem{Liu:2023wfo} Z.~W.~Liu, J.~X.~Lu, M.~Z.~Liu and L.~S.~Geng, Distinguishing the spins of $P_c(4440)$ and $P_c(4457)$ with femtoscopic correlation functions, Phys. Rev. D \textbf{108}, L031503 (2023).
\bibitem{Torres-Rincon:2023qll} J.~M.~Torres-Rincon, {\`A}.~Ramos and L.~Tolos, Femtoscopy of $D$ mesons and light mesons upon unitarized effective field theories, Phys. Rev. D \textbf{108}, 096008 (2023).
\bibitem{Liu:2023huu} X.~Liu, Y.~Tan, D.~Chen, H.~Huang and J.~Ping, Spectroscopy and femtoscopic correlation function of the $B\bar{D}$, $B=(N, \Delta)$ system in quark delocalization color screening model, [arXiv:2307.05516 [hep-ph]].
\bibitem{Albaladejo:2023wmv} M.~Albaladejo, A.~Feijoo, I.~Vida{\~n}a, J.~Nieves and E.~Oset, Inverse problem in femtoscopic correlation functions: the $T_{cc}(3875)^+$ state, Eur. Phys. J. A \textbf{61}, 187 (2025).
\bibitem{Li:2024tof} H.~P.~Li, J.~Y.~Yi, C.~W.~Xiao, D.~L.~Yao, W.~H.~Liang and E.~Oset, Correlation function and the inverse problem in the $BD$ interaction, Chin. Phys. C \textbf{48}, 053107 (2024).
\bibitem{Liu:2024nac} Z.~W.~Liu, J.~X.~Lu, M.~Z.~Liu and L.~S.~Geng, Femtoscopy can tell whether $Z_c(3900)$ and $Z_{cs}(3985)$ are resonances, virtual states, or bound states, Sci. Bull. \textbf{70}, 3515 (2025).
\bibitem{Abreu:2025jqy} L.~M.~Abreu and J.~M.~Torres-Rincon, Searching for femtoscopic signatures of the $D\bar{D}$ ($I$ = 0), $X(3700)$, bound state, Phys. Rev. D \textbf{112}, 016003 (2025).
\bibitem{Liu:2025oar} Z.~W.~Liu, D.~L.~Ge, J.~X.~Lu, M.~Z.~Liu and L.~S.~Geng, Charmonium--nucleon femtoscopic correlation function, Phys. Rev. D \textbf{112}, 054019 (2025).
\bibitem{Etminan:2025tiy} F.~Etminan, Quantitative predictions of alpha-charmonium correlation functions in high-energy collisions, Sci. Rep. \textbf{16}, 20811 (2026).
\bibitem{Barbat:2025orm} M.~F.~Barbat, J.~M.~Torres-Rincon, A.~Ramos and L.~Tolos, Femtoscopy of $DN$ and $\bar{D}N$ systems, Phys. Rev. D \textbf{113}, 056025 (2026).
\bibitem{Agatao:2025ckp} B.~Agat{\~a}o, P.~Brand{\~a}o, A.~Mart{\'\i}nez Torres, K.~P.~Khemchandani, L.~M.~Abreu and E.~Oset, Correlation functions for $n\,\bar{D}_{s1}(2460)$ and $n\,\bar{D}_{s1}(2536)$, Eur. Phys. J. C \textbf{85}, 1136 (2025).
\bibitem{Liu:2025wwx} H.~N.~Liu, Z.~W.~Liu, L.~Abreu and L.~S.~Geng, Traces of the $X(3960)$ state in the femtoscopic $D_{s}^{+}D_{s}^{-}$ correlations, Phys. Rev. D \textbf{113}, 114005 (2026).
\bibitem{Zhang:2025szg} B.~Zhang, Probing the strong interaction between charm hadrons and charged particles with femtoscopy measurements with ALICE, EPJ Web Conf. \textbf{364}, 05006 (2026).
\bibitem{Liu:2025nze} Z.~W.~Liu, J.~M.~Xie, J.~X.~Lu and L.~S.~Geng, Probing the di-$J/\psi$ interaction and the nature of $X(6200)$ with femtoscopic correlation functions, [arXiv:2512.10459 [hep-ph]].
\bibitem{Ge:2026moy} D.~L.~Ge, Z.~W.~Liu and L.~S.~Geng, $DD^*$ correlation functions in deciphering the nature of $T_{cc}(3875)^+$, Phys. Rev. D \textbf{114}, 014054 (2026).
\bibitem{Zhao:2026tpj} J.~Zhao, T.~Song, J.~Aichelin, E.~Bratkovskaya, P.~B.~Gossiaux and K.~Werner, Probe charmonium-nucleon interactions in high energy proton-proton collisions, [arXiv:2603.27391 [hep-ph]].
\bibitem{Shen:2025qpj} Y.~b.~Shen, Z.~W.~Liu, J.~X.~Lu, M.~Z.~Liu and L.~S.~Geng, Probing the structure of the $D_{s 0}^*(2317)$ and $X(3872)$ states through correlation functions, Phys. Lett. B \textbf{878}, 140535 (2026).
\bibitem{Shi:2026hwn} P.~P.~Shi, M.~Albaladejo, F.~K.~Guo and J.~Nieves, A femtoscopic tale of two $C$-parities: the $Z_c(3900)$ and the isovector partner of the $X(3872)$, [arXiv:2608.01237 [hep-ph]].
\bibitem{Song:2026spz} J.~Song, P.~Brandao and E.~Oset, $KX(3872)$ interaction and correlation function, [arXiv:2607.06317 [hep-ph]].

% heavy
\bibitem{Barbat:2026drc} M.~F.~Barbat, J.~Nieves and L.~Tolos, Scattering and Femtoscopic Correlation Functions of the $\Sigma_c^{++}\pi^{+}$, $\Sigma_c^{0}\pi^{-}$ and $\Sigma_b^{+}\pi^{+}$ Systems, Phys. Lett. B \textbf{878}, 140529 (2026).
\bibitem{Feijoo:2023sfe} A.~Feijoo, L.~R.~Dai, L.~M.~Abreu and E.~Oset, Correlation function for the $T_{bb}$ state: Determination of the binding, scattering lengths, effective ranges, and molecular probabilities, Phys. Rev. D \textbf{109}, 016014 (2024).
\bibitem{Lai:2026gql} Z.~T.~Lai, J.~X.~Lu, Z.~W.~Liu, A.~Martinez Torres, K.~P.~Khemchandani and L.~S.~Geng, Revealing the nature of double-strangeness pentaquark states via femtoscopic correlation functions, [arXiv:2606.16727 [hep-ph]].
\bibitem{Liu:2026zlk} S.~W.~Liu, W.~T.~Lyu and J.~J.~Xie, Probing the hadronic molecular nature of the $\Omega(2012)$, $\Omega(2380)$, and $\Omega_c(3120)$ via femtoscopy correlation functions, [arXiv:2604.25630 [hep-ph]].
\bibitem{Jia:2026iqo} W.~H.~Jia, H.~P.~Li, W.~H.~Liang, J.~Song and E.~Oset, Correlation function and bound state from the $K D_{s0}^*(2317)$ interaction, [arXiv:2604.07261 [hep-ph]].
\bibitem{Ikeno:2025bsx} N.~Ikeno and E.~Oset, Correlation function for the $nD^*_{s0}(2317)$ interaction and the issue of elastic unitarity, Phys. Rev. D \textbf{112}, 094019 (2025).

% NLambda_c
\bibitem{Liu:2011xc} Y.~R.~Liu and M.~Oka, $\Lambda_c N$ bound states revisited, Phys. Rev. D \textbf{85}, 014015 (2012).
\bibitem{Maeda:2015hxa} S.~Maeda, M.~Oka, A.~Yokota, E.~Hiyama and Y.~R.~Liu, A model of charmed baryon{\textendash}nucleon potential and two- and three-body bound states with charmed baryon, PTEP \textbf{2016}, 023D02 (2016).
\bibitem{Garcilazo:2019ryw} H.~Garcilazo, A.~Valcarce and T.~F.~Caram{\'e}s, Charmed baryon{\textendash}nucleon interaction, Eur. Phys. J. C \textbf{79}, 598 (2019).
\bibitem{Vidana:2019amb} I.~Vida{\~n}a, A.~Ramos and C.~E.~Jimenez-Tejero, Charmed nuclei within a microscopic many-body approach, Phys. Rev. C \textbf{99}, 045208 (2019).

% LQCD
\bibitem{Miyamoto:2017tjs} T.~Miyamoto, S.~Aoki, T.~Doi, S.~Gongyo, T.~Hatsuda, Y.~Ikeda, T.~Inoue, T.~Iritani, N.~Ishii and D.~Kawai, \textit{et al.}, $\Lambda_c N$ interaction from lattice QCD and its application to $\Lambda_c$ hypernuclei, Nucl. Phys. A \textbf{971}, 113 (2018).
\bibitem{Miyamoto:2017ynx} T.~Miyamoto [HAL QCD], Coupled-channel $\Lambda_c N - \Sigma_c N$ interaction from lattice QCD, PoS \textbf{Hadron2017}, 146 (2018).

% EFT
\bibitem{Haidenbauer:2017dua} J.~Haidenbauer and G.~Krein, Scattering of charmed baryons on nucleons, Eur. Phys. J. A \textbf{54}, 199 (2018).
\bibitem{Song:2020isu} J.~Song, Y.~Xiao, Z.~W.~Liu, C.~X.~Wang, K.~W.~Li and L.~S.~Geng, $\Lambda_cN$ interaction in leading-order covariant chiral effective field theory, Phys. Rev. C \textbf{102}, 065208 (2020).

% pLambda_c CF
\bibitem{Haidenbauer:2020kwo} J.~Haidenbauer, G.~Krein and T.~C.~Peixoto, Femtoscopic correlations and the $\Lambda_c N$ interaction, Eur. Phys. J. A \textbf{56}, 184 (2020).
\bibitem{Zheng:2026qtk} R.~Y.~Zheng, Z.~W.~Liu and L.~S.~Geng, $\Lambda_c N$ correlation functions with leading-order covariant chiral interactions, Phys. Rev. D \textbf{114}, 014006 (2026).

% Charm hypernuclei
\bibitem{Dover:1977jw} C.~B.~Dover and S.~H.~Kahana, Possibility of Charmed Hypernuclei, Phys. Rev. Lett. \textbf{39}, 1506 (1977).
\bibitem{Krein:2017usp} G.~Krein, A.~W.~Thomas and K.~Tsushima, Nuclear-bound quarkonia and heavy-flavor hadrons, Prog. Part. Nucl. Phys. \textbf{100}, 161 (2018).
\bibitem{Hosaka:2016ypm} A.~Hosaka, T.~Hyodo, K.~Sudoh, Y.~Yamaguchi and S.~Yasui, Heavy Hadrons in Nuclear Matter, Prog. Part. Nucl. Phys. \textbf{96}, 88 (2017).
\bibitem{Gibson:1983zw} B.~F.~Gibson, G.~Bhamathi, C.~B.~Dover and D.~R.~Lehman, BINDING ENERGY ESTIMATES FOR CHARMED FEW BODY SYSTEMS, Phys. Rev. C \textbf{27}, 2085 (1983).

% hypernulei
\bibitem{Haidenbauer:2019boi} J.~Haidenbauer, U.~G.~Mei{\ss}ner and A.~Nogga, Eur. Phys. J. A \textbf{56}, 91 (2020).
\bibitem{STAR:2022fnj} B.~Aboona \textit{et al.} [STAR], Phys. Rev. Lett. \textbf{130}, 212301 (2023).
\bibitem{Chen:2023mel} J.~Chen, X.~Dong, Y.~G.~Ma and Z.~Xu, Sci. Bull. \textbf{68}, 3252-3260 (2023).
\bibitem{ALICE:2022sco} S.~Acharya \textit{et al.} [ALICE], Phys. Rev. Lett. \textbf{131}, 102302 (2023).
\bibitem{Ma:2023} Y.~G.~Ma, Hypernuclei as a laboratory to test hyperon–nucleon interactions, Nuc. Sci. Tech. \textbf{34}, 97 (2023).

% Charm hypernuclei
\bibitem{Bando:1981ti} H.~Bando and M.~Bando, $^{5}$He ($\Lambda(c$)) and $^{9}$Be ($\Lambda(c$)) Charmed Nuclei Versus $^{5}$He ($\Lambda$) and $^{9}$Be ($\Lambda$) Hypernuclei, Phys. Lett. B \textbf{109}, 164 (1982).
\bibitem{Haidenbauer:2020uci} J.~Haidenbauer, A.~Nogga and I.~Vida{\~n}a, Predictions for charmed nuclei based on $Y_c N$ forces inferred from lattice QCD simulations, Eur. Phys. J. A \textbf{56}, 195 (2020).
\bibitem{Wu:2020nin} L.~Wu, J.~Hu and H.~Shen, Single $\Lambda_c^+$ hypernuclei within quark mean-field model, Phys. Rev. C \textbf{101}, 024303 (2020).
\bibitem{Liu:2023txn} Y.~X.~Liu, C.~F.~Chen, Q.~B.~Chen, H.~T.~Xue, H.~J.~Schulze and X.~R.~Zhou, Deformed charmed hypernuclei, Phys. Rev. C \textbf{108}, 064312 (2023).
\bibitem{Yang:2024ats} W.~Yang, S.~Y.~Ding and B.~Y.~Sun, Charmed hypernuclei within density-dependent relativistic mean-field theory, Phys. Rev. C \textbf{110}, 054320 (2024).

% previous work
\bibitem{Huang:2013zva} H.~Huang, J.~Ping and F.~Wang, $N \Sigma_c$ and $N \Sigma_b$ resonances in the quark-delocalization color-screening model, Phys. Rev. C \textbf{87}, 034002 (2013).
\bibitem{Wu:2023yux} S.~Wu, Q.~Wu, H.~Huang, X.~Chen, J.~Ping and Q.~Wang, Possibility of generating the $^3_{\Lambda_c}\text{H}$ nucleus in the quark-delocalization color-screening model, Phys. Rev. C \textbf{109}, 014001 (2024).

% QDCSM CF
\bibitem{Yan:2024aap} Y.~Yan, Q.~Huang, Y.~Yang, H.~Huang and J.~Ping, Investigating the $p$--$\Omega$ interactions and correlation functions, Sci. China Phys. Mech. Astron. \textbf{68}, 232012 (2025).
\bibitem{Yan:2025hpa} Y.~Yan, Q.~Huang, Q.~Wu, H.~Huang and J.~Ping, Prediction of $p\bar{\Omega }$ states and femtoscopic study, Nucl. Sci. Tech. \textbf{37}, 126 (2026).
\bibitem{Yan:2026yrd} Y.~Yan, Y.~Wu, Y.~Tan, X.~Hu, Q.~Huang, H.~Huang and J.~Ping, Investigating the $\Omega$--$\phi$ interaction and correlation functions, Phys. Rev. C \textbf{113}, 065204 (2026).

% QDCSM
\bibitem{Wang:1992wi} F.~Wang, G.~h.~Wu, L.~j.~Teng and J.~T.~Goldman, Quark delocalization, color screening, and nuclear intermediate range attraction, Phys. Rev. Lett. \textbf{69}, 2901 (1992).
\bibitem{Wu:1998wu} G.~h.~Wu, J.~L.~Ping, L.~j.~Teng, F.~Wang and J.~T.~Goldman, Quark delocalization, color screening model and nucleon baryon scattering, Nucl. Phys. A {\bf 673}, 279 (2000).
\bibitem{Pang:2001xx} H.~R.~Pang, J.~L.~Ping, F.~Wang and J.~T.~Goldman, Phenomenological study of hadron interaction models, Phys. Rev. C {\bf 65}, 014003 (2002).

% Framework
\bibitem{ParticleDataGroup:2026aaa} F.~Takahashi \textit{et al.} [Particle Data Group], Review of Particle Physics, Int. J. Mod. Phys. A 41, 2630011 (2026).
\bibitem{Xu:2007oam} M.~Xu, M.~Yu and L.~Liu, Examining the crossover from hadronic to partonic phase in QCD, Phys. Rev. Lett. \textbf{100}, 092301 (2008).
\bibitem{Yan:2024usf} Y.~Yan, Q.~Huang, X.~Zhu, H.~Huang and J.~Ping, Investigating $\Xi$ resonances from a pentaquark perspective, Phys. Rev. D \textbf{110}, 014021 (2024).

% KP
\bibitem{Pratt:1990zq} S.~Pratt, T.~Csorgo and J.~Zimanyi, Detailed predictions for two pion correlations in ultrarelativistic heavy ion collisions, Phys. Rev. C \textbf{42}, 2646 (1990).
\bibitem{Bauer:1992ffu} W.~Bauer, C.~K.~Gelbke and S.~Pratt, Hadronic interferometry in heavy ion collisions, Ann. Rev. Nucl. Part. Sci. \textbf{42}, 77 (1992).
% CATS
\bibitem{Mihaylov:2018rva} D.~L.~Mihaylov, V.~Mantovani Sarti, O.~W.~Arnold, L.~Fabbietti, B.~Hohlweger and A.~M.~Mathis, A femtoscopic Correlation Analysis Tool using the Schrödinger equation (CATS), Eur. Phys. J. C \textbf{78}, 394 (2018).

% RGM
\bibitem{Wheeler:1937zza} J.~A.~Wheeler, Molecular Viewpoints in Nuclear Structure, Phys. Rev. \textbf{52}, 1083 (1937).
\bibitem{Yan:2023tvl} Y.~Yan, X.~Hu, H.~Huang and J.~Ping, Investigating excited $\Omega_c$ states from pentaquark perspective, Phys. Rev. D \textbf{108}, 094045 (2023).

% KHK
\bibitem{Kamimura:1977okl} M.~Kamimura, Chapter V. A Coupled Channel Variational Method for Microscopic Study of Reactions between Complex Nuclei, Prog. Theor. Phys. Suppl. \textbf{62}, 236 (1977).

% GEM
\bibitem{Hiyama:2003cu} E.~Hiyama, Y.~Kino and M.~Kamimura, Gaussian expansion method for few-body systems, Prog. Part. Nucl. Phys. \textbf{51}, 223 (2003).
\bibitem{Hiyama:2019kpw} E.~Hiyama, K.~Sasaki, T.~Miyamoto, T.~Doi, T.~Hatsuda, Y.~Yamamoto and T.~A.~Rijken, Possible lightest $\Xi$ Hypernucleus with Modern $\Xi N$ Interactions, Phys. Rev. Lett. \textbf{124}, 092501 (2020).

% emission source
\bibitem{ALICE:2023sjd} S.~Acharya \textit{et al.} [ALICE], Common femtoscopic hadron-emission source in $pp$ collisions at the LHC, Eur. Phys. J. C \textbf{85}, 198 (2025).
\bibitem{Mihaylov:2023pyl} D.~Mihaylov and J.~Gonz{\'a}lez Gonz{\'a}lez, Novel model for particle emission in small collision systems, Eur. Phys. J. C \textbf{83}, 590 (2023).
\bibitem{VazquezDoce:2024nye} O.~V{\'a}zquez Doce, D.~Mihaylov and L.~Fabbietti, Study of the deuterons emission time in $pp$ collisions at the LHC via kaon-deuteron correlations, Eur. Phys. J. A \textbf{61}, 53 (2025).
\bibitem{Wang:2025htd} D.~F.~Wang, M.~Y.~Chen, Y.~G.~Ma, Q.~Y.~Shou, S.~Zhang and L.~Zheng, Investigating the pion emission source in $pp$ collisions using the AMPT model with subnucleon structure, Nucl. Sci. Tech. \textbf{36}, 154 (2025).








\end{thebibliography}
\end{document}